\documentstyle[preprint,aps]{revtex}
\begin{document}
\preprint{\sl Int. J. Mod. Phys. A (in press)}
\hsize = 7.0in
\widetext
\draft
\tighten
\title{Theory of Neutral Particles:
McLennan-Case Construct for Neutrino, its Generalization, and
a  New Wave Equation}

\author{D. V. Ahluwalia}

\address {Mail Stop H 846, Group P-25\\
Los Alamos Meson Physics Facility\\
Los Alamos National Laboratory,  Los Alamos, New Mexico 87545, USA \\
Internet address:   AV@LAMPF.LANL.GOV}

\maketitle

\begin{abstract}

Continuing our recent argument where we constructed a FNBWW-type spin-$1$
boson having opposite relative intrinsic parity to that of the associated
antiparticle, we now study eigenstates of the Charge Conjugation operator.
Based on the observation that if $\phi_{_{L}}(p^\mu)$ transforms as a
$(0,\,j)$ spinor under Lorentz boosts, then
$\Theta_{[j]}\,\phi_{_{L}}^\ast(p^\mu)$ transforms as a $(j,\,0)$ spinor
(with a similar relationship existing between $\phi_{_{R}}(p^\mu)$ and
$\Theta_{[j]}\,\phi_{_{R}}^\ast(p^\mu)$; where $ \Theta_{[j]}\,{\bf
J}\,\Theta_{[j]}^{-1}\,=\,-\,{\bf J}^\ast $ with $\Theta_{[j]}$ the well
known Wigner matrix involved in the operation of time reversal) we
introduce McLennan-Case type $(j,\,0)\oplus(0,\,j)$ spinors. Relative phases
between $\phi_{_{R}}(p^\mu)$ and $\Theta_{[j]}\,\phi_{_{R}}^\ast(p^\mu)$,
and $\Theta_{[j]}\,\phi_{_{L}}^\ast(p^\mu)$ and $\phi_{_{L}}(p^\mu)$, turn
out to have physical significance and are fixed by appropriate
requirements. Explicit construction, and a series of physically relevant
properties, for these spinors are obtained for spin-$1/2$ and spin-$1$
culminating in the construction of a new wave equation and
introduction of Dirac-like and Majorana-like quantum fields.

\end{abstract}


\vskip 0.3in

While in  the case of gravitation the dynamical role played by space-time
symmetries is manifest, the dynamical role played by space-time symmetries
for other interactions is not always fully appreciated. This was recently
emphasized in Ref \cite{MSb}. Consider, as an example, quantum
electrodynamics (QED). An essential element of the QED Lagrangian is the
Dirac operator $(i\,\gamma^\mu\,\partial_\mu\,-\,m\,\openone)$ for the
charged fermion field $\Psi(x)$. The demand of covariance under the local
gauge transformation, $\Psi(x)\rightarrow
\exp{\bbox(}i\,\alpha(x){\bbox)}\,\Psi(x)$, introduces a local $U(1)$
interaction via the   vector potential $A_\mu(x)$. The basic
building block, i.e., the operator
$(i\,\gamma^\mu\,\partial_\mu\,-\,m\,\openone)$, of the QED Lagrangian
follows
directly from the space-time symmetries
and it contains significant
information regarding the dynamical behavior and kinematical properties, such
as relative intrinsic parities of the charged
fermions and antifermions of spin-$1/2$, of
the system. \footnotemark[1]
\footnotetext[1]
{
To see how simply and deeply
is the operator
$(i\,\gamma^\mu\,\partial_\mu\,-\,m\,\openone)$ connected with the
space-time symmetries,
the reader may first refer to Sec. 2.3 of Ryder's
book \cite{LHR}, next study Ref. \cite{FNBWW}, and finally refer to the
text bracketed between Eqs. (1) and (17) of Ref. \cite{Vorticity}, which
provides some of the missing details of the previous two references.
Here is the argument in brief: Refer to Eqs. (\ref{r}) and
(\ref{l}), and the surrounding definitions,
of this paper and set $\bbox{J}\,=\,\bbox{\sigma}/2$.
Next, note that spinors [implied by
the arguments based on  parity symmetry and that Lorentz group is
essentially $SU_{_R}(2)\otimes SU_{_L}(2)$]
\[
\psi(p^\mu)\,\equiv\,
\left(
\begin{array}{c}
\phi_{_R}(p^\mu)\\
\phi_{_L}(p^\mu)
\end{array}\right)
\]
turn out to be of crucial significance in constructing a field
$\Psi(x)$ that describes eigenstates of the Charge operator, $Q$, {\it if}
$\phi_{_R}(\overcirc{p}^\mu)\,=\,\pm\,\phi_{_L}(\overcirc{p}^\mu)$
(otherwise physical eigenstates are no longer charge eigenstates).
We call $\phi_{_R}(\overcirc{p}^\mu)\,=\,\pm\,\phi_{_L}(\overcirc{p}^\mu)$,
the ``Ryder-Burgard relation'' (see Ref. \cite{LHR} and footnote [11]).
Next couple
the Ryder-Burgard relation with
Eqs.
 (\ref{r}) and
(\ref{l})
to obtain (useful identities: $\cosh(\bbox{\sigma\cdot\varphi})\,=\,
\openone\,\cosh\varphi$,
$\sinh(\bbox{\sigma\cdot\varphi})\,=\,\bbox{\sigma\cdot\widehat{p}}\,
\sinh\varphi$,  $\cosh\varphi\,=\,E/m$,  $\sinh\varphi\,=\,
\vert\bbox{p}\vert/m)$
\[
\left(\begin{array}{cc}
\mp\,m\,\openone & p_0\,+\,\bbox{\sigma\cdot p} \\
p_0\,-\,\bbox{\sigma\cdot p} &\mp\,m\,\openone
\end{array}\right)\,\psi(p^\mu)\,=\,0\quad.
\]
Introducing $\psi(x)\,\equiv\,\psi(p^\mu)\,\exp(\mp\,i p\cdot x)$ and letting
$p_\mu\,\rightarrow \,i\,\partial_\mu$, the above equation becomes:
$(i\,\gamma^\mu\,\partial_\mu\,-\,m\,\openone)\psi(x)\,=\,0\,$.
This is the Dirac equation for spin-$1/2$ particles with $\gamma^\mu$ in
the Weyl/Chiral representation. Similarly, one can
obtain wave equations and thus a complete kinematic structure
and the associated dynamical consequences for other
Dirac-like $(j,\,0)\oplus(0,\,j)$ spinors $\psi(p^\mu)$ and quantum fields
$\Psi(x)$. Also see footnote 7.}
Without this, or similar, kinematical
structure we would not even know how to formulate a principle of local
gauge invariance for QED. Like mass and spin,
\footnotemark[2]
\footnotetext[2]{Recall that
mass and spin are intimately related with the two
Casimir invariants of the Poincar\'e group. The two Casimir invariants are
$C_1\,\equiv\, P^\mu\,P_\mu$, and $C_2\,\equiv\,W^\mu\,W_\mu$ with the
Pauli-Lubanski pseudovector defined as  $W_\mu\,\equiv\,- (1/2)\,
\epsilon_{\mu\nu\rho\sigma}\,J^{\nu\rho}\,P^\sigma$ \cite{EPW}.}
 the concept of charge conjugation
\footnotemark[3]\footnotetext[3]
{The positive- and negative-energy solutions associated with the
Dirac equation purely at the kinematical level essentially suffice to
suggest the concept of Charge Conjugation.} and the associated notion of
antiparticles arises from the Poincar\'e space-time symmetries.

The fact
that QED, and presumably QCD, fermions are found in charge eigenstates is
deeply connected with the structure of Dirac's $(1/2,\,0)\oplus(0,\,1/2)$
representation space. In fact, parity covariance is built into QED by
building the kinematical structure of the theory on  the Dirac's
$(1/2,\,0)\oplus(0,\,1/2)$ field rather than the Weyl's $(1/2,\,0)$ or
$(0,\,1/2)$ field. From Majorana's work \cite{EM} one knows that Dirac's
construct is not the only construct possible for the
$(1/2,\,0)\oplus(0,\,1/2)$ quantum field;
moreover, we shall see that there are additional possibilities.
An important reformulation  of Majorana's work was undertaken by
McLennan and Case \cite{JAM,KMC} in 1957. Here we shall
extend the McLennan-Case formulation and
generalize it to spin-$1$ (and
higher). While considering spin-$1$, a surprising conclusion is
reached: Within the framework of the formalism developed in this paper,
there are no self-charge conjugate spinors in the $(1,\,0)\oplus(0,\,1)$
representation space. Fundamentally new wave equations, non-unitarily
connected to Dirac (for spin-$1/2$) or modified Weinberg equation (for
spin-$1$), will be presented.

Some of the constructs presented may appear familiar and well
known. For example,  half of the  type II $(1/2,0)\oplus(0,1/2)$
spinors presented
below are identical with  Majorana spinors (Ref. \cite{PR}, p. 20)
The remaining half of the type II spinors for spin-$1/2$, introduced as
anti-self charge conjugate counterparts of Majorana spinors, are needed
to obtain a complete set (``complete'' in the mathematical sense).
The extension to higher spins is accomplished by recognizing that the
``magic of the Pauli matrices'', Ref. \cite{PR} Eq. 1.4.24, is in fact
a special example of the relation
$
\Theta_{[j]}\,{\bf J}\,\Theta_{[j]}^{-1}\,=\,-\,{\bf J}^\ast
$
with $\Theta_{[j]}$ the well known Wigner matrix involved in the
operation of time reversal.
To my knowledge, no equation of motion for these
spinors in
their $2(2j+1)$-element form exists in literature.
This fact in my view has prevented the full exploitation, and understanding
of the precise physical content, of a field theory based on Majorana
(and the associated anti-self charge conjugate spinors) spinors. It is
hoped that the fundamentally new wave equation
derived
in this work will remedy the existing situation and lead to a
deeper understanding of the work initiated by
McLennan and Case.
\footnotemark[4]
\footnotetext[4]
{
The Jehle \cite{HJ} type
equations that appear in Refs. \cite{JAM,KMC} for spin-1/2 (and its simple
generalization to higher spins) while consistent with our work, are far from
being wave equations for the $2(2j+1)$-element $(j,\,0)\oplus(0,\,j)$ spinors
of
type-II introduced in this work. The exact meaning of this remark will become
clear as we proceed.}
Also the facts that the type II spinors cannot be helicity eigenspinors, or
that their properties are unusual under the operation of parity, or that
introducing gauge interactions via the standard minimal substitution for such
particles is impossible, are, to the best of my knowledge, all new results.

It is hoped that the formalism developed here will provide, if so
necessitated by future theoretical and (or) experimental reasons,  a
from-the-first-principle point of departure for  the
kinematical and dynamical understanding of neutrinos and photinos. For
instance, if the phenomenon of neutrino oscillations is
established \cite{LSND} then
one will be forced to express the weak eigenstates as a linear
superposition of mass eigenstates. In this context the question would
arise if the mass eigenstates can have a description beyond the well known
Dirac and Majorana states and on what theoretical/experimental grounds one
will choose the various mass eigenstates in the superposition. Experimental
results, in general, cannot be insensitive to such choices and the P, CP,
transformation of the states will depend on the choices made in the
theoretical models.

To establish the setting of the ideas it is perhaps appropriate to orient
the discussion by making a few brief observations on the subject.
It was in
1927 that Wigner introduced the notion of
parity  in quantum mechanics \cite{EPW_P}. In his celebrated paper of 1939
\cite{EPW} and his 1962 collaborative work with Bargmann and Wightman,
without reference to any wave equations, Wigner \cite{BWEPW} classified
quantum field theories on the basis of their behavior under continuous
space-time symmetries and the operations of discrete symmetries of Parity,
Time Reversal and Charge Conjugation. Since
the transformation properties of quantum mechanical states do not directly
invoke a wave equation and since the operation of Charge Conjugation cannot
be determined without exploiting an appropriate wave equation,  a specific
connection between the Lorentz group representations and their behavior
under the operations of Parity, Time Reversal and Charge Conjugation had
remained essentially unexplored beyond spin-$1/2\,$. While Weinberg
\cite{SW} did undertake a general study of this connection in 1964 for the
$(j,\,0)\oplus(0,\,j)$ representation space,  various details escaped
notice until a recent publication \cite{FNBWW}. In Ref. \cite{FNBWW} it was
shown that the $(1,\,0)\oplus(0,\,1)$ representation space is a concrete
realization of a FNBWW (Foldy-Nigam-Bargmann-Wightman-Wigner
\footnotemark[5] \footnotetext[5]
{After the publication of Ref.\cite{FNBWW}, and after much of the present
work was already in draft form,  we learned that ideas similar to those  of
Bargmann, Wightman, and Wigner were put forward previously and
independently by Foldy and Nigam \cite{F,NF}. I thank Dr. Zurab K.
Silagadze for bringing Foldy's paper to my attention. Professor Foldy
brought Ref. \cite{NF} to my attention where, in essence, Nigam and Foldy
constructed a quantum field theory where a spin-$1/2$ fermion and its
antifermion have same relative intrinsic parity. It is my pleasure to thank
Professors Foldy and Nigam for  conversations and correspondence.})
type quantum
field theory in which a boson and its antiboson carry {\it opposite}
relative intrinsic parities. In Ref. \cite{FNBWW} we confined our attention
to particles that are eigenstates of the Charge operator. In this paper, by
going beyond the eigenstates of the Charge operator,
I provide a further instance of establishing the kinematical structure of a
field theory upon the relevant space-time symmetries.

To avoid possible confusion,  we parenthtically  note that the
representation space under consideration in Weinberg's work \cite{SW} and
the present work is the $(j,0)\oplus(0,j)$ representation space. Contrary
to the canonical wisdom,  we have recently established \cite{FNBWW} that
certain fundamental aspects of a quantum field theory are deeply connected
with the representation space one chooses to describe a spin-$j$ particle.
For description of particles outside the $(j,0)\oplus(0,j)$ representation
space there are well known works and some of the fundamental papers are
catlogued in Ref. \cite{Proca}

\bigskip

\noindent {\it Two Types of $(j,\,0)\oplus(0,\,j)$ Spinors}

To define our conventions and notation, following the 1939 classic work of
Wigner \cite{EPW} we note that, without reference to any wave equation, the
$(j,\,0)$ and $(0,\,j)$ spinors, in the notation of our earlier work
\cite{FNBWW}, under {\it Lorentz} transformations transform as
\begin{mathletters} \begin{eqnarray} &&(j,\,0):\quad \phi_{_R}( p^\mu)
\,=\, \Lambda_{_R}(p^\mu\leftarrow\overcirc{p}^\mu)
\,\phi_{_R}(\overcirc{p}^\mu)\quad,\label{r}\\
&&(0,\,j):\quad\phi_{_L}(p^\mu)\,
\,=\,\Lambda_{_L}(p^\mu\leftarrow\overcirc{p}^\mu)\,
\phi_{_L}(\overcirc{p}^\mu)\quad. \label{l} \end{eqnarray}
\end{mathletters} In the above expressions   $\overcirc{p}^\mu$ refers to
the energy-momentum four vector associated with the particle, of mass $m$,
at rest;
and  the boost parameter $\bbox{\varphi}$
is defined as in Eq. (3) of Ref. \cite{FNBWW}.
The Wigner boosts for the $(j,\,0)$ and  $(0,\,j)$ spinors are
\begin{equation}
\Lambda_{_R}(p^\mu\leftarrow\overcirc{p}^\mu)
\,=\,\exp\left(i\,{\bf K^{(j,\,0)}\,\cdot}\bbox{\varphi}\right)\,\,,\quad
\Lambda_{_L}(p^\mu\leftarrow\overcirc{p}^\mu)\,=\,
\exp\left(i\,{\bf K^{(0,\,j)}\,\cdot}\bbox{\varphi}\right)
\quad. \label{wrl}
\end{equation}
For the $(j,\,0)$ representation space ${\bf B}\,\equiv\,2^{-1}
({\bf J}\,-\,i\,{\bf K})$ equals zero, and therefore ${\bf K^{(j,\,0)}}\,=\,
-\,i\,{\bf J}$.
Similarly,  for the $(0,\,j)$ representation space ${\bf A}\,\equiv\,2^{-1}
({\bf J}\,+\,i\,{\bf K})$ equals zero, and therefore ${\bf K^{(0,\,j)}}\,=\,
+\,i\,{\bf J}$. Consequently,
\begin{equation}
\Lambda_{_R}(p^\mu\leftarrow\overcirc{p}^\mu)
\,=\,\exp\left({\bf J}\,\cdot\bbox{\varphi}\right)\label{wjr}\,\,,\quad
\Lambda_{_L}(p^\mu\leftarrow\overcirc{p}^\mu)\,=\,
\exp\left(-\,{\bf J}\,\cdot\bbox{\varphi}\right)
\quad. \label{wjrl}
\end{equation}
Here, ${\bf J}$ are the standard spin-$j$ matrices with $J_z$
diagonal.
Having established our conventions and notation, we now note that
$
\left[\Lambda_{_{L,R}}(p^\mu\leftarrow\overcirc{p}^\mu)\right]^{-1}
\,=\,
\left[\Lambda_{_{R,L}}(p^\mu\leftarrow\overcirc{p}^\mu)\right]^\dagger
$ and the fact that Wigner's  operator  $\Theta_{[j]}$
for spin-$j$ has the property
$
\Theta_{[j]}\,{\bf J}\,\Theta_{[j]}^{-1}\,=\,-\,{\bf J}^\ast
$
implies that if $\phi_{_{L}}(p^\mu)$ transforms as a $(0,\,j)$ spinor
under Lorentz boosts,
then $\Theta_{[j]}\,\phi_{_{L}}^\ast(p^\mu)$ transforms as a $(j,\,0)$
spinor.
Similarly,
if $\phi_{_{R}}(p^\mu)$ transforms as a $(j,\,0)$ spinor under Lorentz
boosts, then $\Theta_{[j]}\,\phi_{_{R}}^\ast(p^\mu)$ transforms as a
$(0,\,j)$ spinor. Wigner's  operator is defined as
$\left(\Theta_{[j]}\right)_
{\sigma,\,\sigma^\prime}\,=\,(-\,1)^{j+\sigma}\,
\delta_{\sigma^\prime,\,-\,\sigma}$ with $\sigma$ and $\sigma^\prime$ as
eigenvalues of $\bf J$. By definition $\Theta_{[j]}$ is real. For
spin-$1/2$ and spin-$1$ it reads:
\begin{equation}
\Theta_{[1/2]}\,=\,\left(\begin{array}{cc}
0&-\,1\\
1&0\end{array}\right)\,\,,\quad
\Theta_{[1]}\,=\,\left(\begin{array}{ccc}
0&0&1\\
0&-\,1&0\\
1&0&0\end{array}\right)\quad.
\end{equation}

These observations allow us to introduce
two types of $(j,\,0)\oplus(0,\,j)$
spinors. The first  type,
\begin{equation}
\psi(p^\mu)\,\equiv\,
\left(
\begin{array}{c}
\phi_{_R}(p^\mu)\\
\phi_{_L}(p^\mu)
\end{array}\right)\quad,\label{ds}
\end{equation}
to be referred as spinors of Type I.
The second type,
\begin{equation}
\lambda(p^\mu)\,\equiv
\left(
\begin{array}{c}
\left(\zeta_\lambda\,\Theta_{[j]}\right)\,\phi^\ast_{_L}(p^\mu)\\
\phi_{_L}(p^\mu)\\
\end{array}
\right)\,\,,\quad
\rho(p^\mu)\,\equiv
\left(
\begin{array}{c}
\phi_{_R}(p^\mu)\\
\left(\zeta_\rho\,\Theta_{[j]}\right)^\ast\,\phi^\ast_{_R}(p^\mu)
\end{array}
\right)
\,\,\quad,\label{os}
\end{equation}
to be referred as spinors of Type II. In Eq. (\ref{os})
$\zeta_\lambda$ and $\zeta_\rho$ are phase factors that
are yet to be fixed.

We now begin with a general study of some of the properties of these
spinors and quantum field theories based on these spinors. The arguments
that follow can be made for any spin, but since spin-$1/2$ and spin-$1$ are
of special prominence in phenomenological descriptions, we confine
ourselves
to these two spins. For spin-$1/2$ the operation of charge conjugation and
parity (``space inversion''
\footnotemark[6]
\footnotetext[6]
{As usual, the operation of parity in space-time is defined as
$
{\cal P}\,:\quad {\bf x}\,\rightarrow\,-\,{\bf x}
\,\,\mbox{and}\,\, t\,\rightarrow\,t\,
$.}
)
on the $(1/2,\,0)\oplus(0,\,1/2)$-representation-space spinors
is given by
\cite{PR,ON}
\begin{equation}
S^c_{[1/2]}
\,=\,
e^{i\vartheta^c_{[1/2]}}
\,
\left(
\begin{array}{cc}
0 & i\,\Theta_{[1/2]}\\
-\,i\,\Theta_{[1/2]} &0
\end{array}\right)\,{\cal K}\,\equiv\,{\cal C}_{[1/2]}\,{\cal K} \,\,,\quad
S^s_{[1/2]}
\,=\,
e^{i\vartheta^s_{[1/2]}}\,
\left(
\begin{array}{cc}
0&\openone_2\\
\openone_2&0
\end{array}
\right)\,=\,e^{i\vartheta^s_{[1/2]}}\,\gamma^{0}
\quad, \label{cph}
\end{equation}
where ${\cal K}$ is the operation of complex conjugation,
and
$\openone_2$ is a $2\times 2$ identity matrix. While this result is well
known, the correct counterpart of results (\ref{cph}) for the
$(1,\,0)\oplus(0,\,1)$ representation space required a careful study. A
detailed analysis was recently presented in Ref. \cite{FNBWW}; here we simply
need the result. These results read:
\begin{equation}
S^c_{[1]}
\,=\,
e^{i\vartheta^{c}_{[1]}}
\,
\left(
\begin{array}{cc}
0 & \Theta_{[1]}\\
-\,\Theta_{[1]} &0
\end{array}\right)\,{\cal K}\,\equiv\,{\cal C}_{[1]}\,{\cal K}\, \,\,,\quad
S^s_{[1]}
\,=\,
e^{i\vartheta^{s}_{[1]}}\,
\left(
\begin{array}{cc}
0&\openone_3\\
\openone_3&0
\end{array}
\right)\,=\,e^{i\vartheta^{s}_{[1]}}\,\gamma_{00}
\quad. \label{cpo}
\end{equation}
Note that neither $S^c_{[1/2]}$ nor  $S^c_{[1]}$ are unitary (or even
linear). This fact, besides others, necessitates   building a quantum field
theory (the so-called ``second quantized theory'') containing a field
operator, $\Psi_{[j]}(x)$ expanded in terms of the above spinors. The
operation of charge conjugation is then defined via a {\it unitary
operator} $U^c_{[j]}$ such that \cite{NF}
\begin{equation}
U^c_{[j]}\,\Psi_{[j]}(x) \Big(U^{c}_{[j]}\Big)^{-1}\,=\,{\cal C}_{[j]}
\Psi^\dagger_{[j]}(x) \quad. \label{qf}
\end{equation}

 For Majorana
fields, internal consistency of the theory requires $\vartheta^s$ and
$\vartheta^c$ to be constrained. However, incorporating such constrained
phase factors can be done as needed and we shall choose $\vartheta^c\,=\,0$
so as to stay as close as possible to Ramond's \cite{PR} discussion on the
subject. A further note of caution is in order. Using the type-I spinors,
one may construct a Majorana field operator for any spin
\cite{IJMPE}. In this formalism the field describes
states of a particle that are simultaneously
eigenstates of the Parity operator (with imaginary eigenvalues
\cite{GR}) and the
Charge Conjugation operator. But how can that be? For spin-$1/2$, for
instance, the usual text-book expressions for
$U^s$ and $U^c$ do not commute! This apparent
inconsistency is resolved  in the careful analysis of the
$U^s$ and $U^c$ operators by Nigam and Foldy \cite{NF} which
reveals that $U^s$
is in fact a function of the Charge operator. For self-charge conjugate
fields, if one makes the additional assumption that the Charge operator
acting on a self-charge-conjugate state yields zero (an assumption certainly
{\it not} valid for charged particles, because then simultaneous
eigenstates of the Charge Conjugation operator and Charge operator are not
possible), one finds that the commutator of the $U^s$ and $U^c$ vanishes
upon
acting on Majorana states. The commutativity/anticommutativity of $U^s$ and
$U^c$ operators is a function of the Charge operator with formal and
phenomenological consequences.

When one constructs a quantum field based on the $(1,\,0)\oplus(0,\,1)$
spinors of  type I,  one finds that the Fock space
operators for charge eigenstates  that
determine transformation of physical states under the operation of Charge
Conjugation, $U^c_{[1]}$, and Parity, $U^s_{[1]}$, {\it anticommute}.
This results in a
quantum field theory of spin-$1$ bosons where a boson and its anti-boson
carry {\it opposite} relative intrinsic parity. For this aspect of the
kinematic structure we refer the reader to our recent publication
\cite{FNBWW}. In the rest of the
paper we confine our attention to spinors of type II,
and the quantum field theory that is built upon them.

{\it We begin with the question: What physical requirement can fix the
phase factors $\zeta$ that appear in the definition of the spinors of Type
II$\,$?} We do not, at the moment, investigate all possible physically
relevant requirements but confine to the most obvious ones.

For spin-$1/2$, we find that the requirement of self/anti-self  conjugacy
under the operation of charge conjugation \begin{equation} S^c_{[1/2]}
\,\lambda(p^\mu)\,=\,\pm\,\lambda(p^\mu)\,\,,\quad S^c_{[1/2]}
\,\rho(p^\mu)\,=\,\pm\,\rho(p^\mu)\quad,\label{sac} \end{equation}
determines $\zeta^S_\lambda\,=\,+\,i\,=\,\zeta^S_\rho$ for the self charge
conjugate spinors $\lambda^S(p^\mu)$ and
$\rho^S(p^\mu)$; and
$\zeta^A_\lambda\,=\,-\,i\,=\,\zeta^A_\rho$ for the anti-self charge conjugate
spinors $\lambda^A(p^\mu)$ and $\rho^A(p^\mu)$. The $\lambda^S(p^\mu)$  are
thus seen to coincide with the McLennan-Case construct \cite{JAM,KMC}.

For spin-$1$, on the other hand, the requirement of self/anti-self charge
conjugacy {\it cannot} be satisfied. That is, there does not exist a
$\zeta$ that can satisfy the spin-$1$ counterpart of the requirement
(\ref{sac}). We find, however, that the requirement of self/anti-self
conjugacy under charge conjugation can be replaced by the requirement
of self/anti-self
conjugacy under the operation of $\Gamma^5\,S^c_{[1]}$, where
$\Gamma^5$ is the {\it chirality} operator for the
$(1,\,0)\oplus(0,\,1)$ representation space and reads:
\footnotemark[7]
\footnotetext[7]{All explicit expressions for the
$2(2j+1)\times 2(2j+1)$ matrices, $\Gamma$, that appear in this
paper are given in the generalized chiral/Weyl (W) representation.
These matrices are related to the generalized canonical/Dirac (D)
representation, which for spin-$1/2$ reduces to the familiar representation
of Bjorken and Drell's text \cite{BD}, via the following
expression
\[
\Gamma_D\,=\,S\,\Gamma_W\,S^{-1}\quad,\qquad
S\,=\,{1\over{\sqrt 2}}\left(
\begin{array}{cc}
\openone &\openone\\
\openone&-\,\openone\end{array}\right)\quad,
\]
with $\openone\,=\,\mbox{a}\,\,(2j+1)\times(2j+1)$ identity matrix.}

\begin{equation}
\Gamma^5\,=\,\left(\begin{array}{cc}
\openone_3&0\\
0&-\openone_3\end{array}\right)\quad,
\label{gf} \end{equation}
with similar expressions for other spins.
The requirement
\begin{equation}
\left[\Gamma^5\,S^c_{[1]}\right]\,
\lambda(p^\mu)\,=\,\pm\,\lambda(p^\mu)\,\,,\quad
\left[\Gamma^5\,S^c_{[1]}\right]\,
\rho(p^\mu)\,=\,\pm\,\rho(p^\mu)\quad,\label{sacc}
\end{equation}
determines $\zeta^S_\lambda\,=\,+\,1\,=\,\zeta^S_\rho$ for the self
$\left[\Gamma^5\,S^c_{[1]}\right]$-conjugate  spinors $\lambda^S(p^\mu)$ and
$\rho^S(p^\mu)$; and
$\zeta^A_\lambda\,=\,-\,1\,=\,\zeta^A_\rho$ for the anti-self
$\left[\Gamma^5\,S^c_{[1]}\right]$-conjugate
spinors $\lambda^A(p^\mu)$ and $\rho^A(p^\mu)$.

For spin-$1/2$, the often repeated assertion that a Majorana spinor is a
Weyl spinor in four-component form is somewhat misleading as a simple
counting of the degrees of freedom immediately reveals. For the massive
case, there are {\it four} $\lambda(p^\mu)$-type spinors -- two
$\lambda^S(p^\mu)$ and two $\lambda^S(p^\mu)$. For comparison, the
associated Weyl spinor, $\phi_{_L}(p^\mu)$,  has only {\it two} degrees of
freedom. Same arguments apply to $\rho(p^\mu)$ and $\phi_{_R}(p^\mu)$. It
will be shown later that $\lambda^S(p^\mu)$ are positive energy solutions
and $\lambda^A(p^\mu)$ are negative energy solutions of an appropriate wave
equation.

\bigskip
\noindent
{\it Helicity and Type-II Spinors}
\bigskip

For convenience, we define self/anti-self $\theta$-conjugacy to be
self/anti-self conjugacy under the operation of charge conjugation for
spin-$1/2$ and under the operation of $\left[\Gamma^5\,S^c_{[1]}\right]$
for spin-1. Let $\phi^{h'}_{_{L,R}}({p}^\mu)$ be an eigenstate of the
helicity operator
\begin{equation}
{\bf J}\cdot\widehat{\bf p}\,\,
\phi^{h'}_{_{L,R}}({p}^\mu)\,=\,h'\,\, \phi^{h'}_{_{L,R}}({p}^\mu)\quad;
\label{ida}
\end{equation}
the Wigner-identity
$
\Theta_{[j]}\,{\bf J}\,\Theta_{[j]}^{-1}\,=\,-\,{\bf J}^\ast
$
then implies that
\begin{equation}
{\bf J}\cdot\widehat{\bf p}\,\,
\Theta_{[j]}\,\left[\phi^{h'}_{_{L,R}}({p}^\mu)\right]^\ast\,=\,-\,h'\,\,
\Theta_{[j]}\,\left[\phi^{h'}_{_{L,R}}({p}^\mu)\right]^\ast
\quad.\label{idb}
\end{equation}
That is, if $\phi^{h'}_{_{L,R}}({p}^\mu)$ are eigenvectors of ${\bf J}
\cdot\widehat{\bf p}$, then
$\Theta_{[j]}\,\left[\phi^{h'}_{_{L,R}}({p}^\mu)\right]^\ast$ are
eigenvectors of ${\bf J} \cdot\widehat{\bf p}$ with {\it opposite}
eigenvalues to those associated with
$\phi^{h'}_{_{L,R}}({p}^\mu)$. An inspection of (\ref{os}) when coupled with
this result implies that
{\it the self/anti-self  $\theta$-conjugate
spinors cannot be in helicity eigenstates.}
The helicity operator for the $(j,\,0)\oplus(0,\,j)$ representation space is
defined as
\begin{equation}
{h}\,\equiv\,\left(\begin{array}{cc}
{\bf J}\cdot\widehat{\bf p} & 0 \\
0&{\bf J}\cdot\widehat{\bf p}
\end{array}\right)\quad.
\end{equation}
Given
$\phi^{h'}_{_{L,R}}({p}^\mu)$ as eigenstates of $ h$, the $\lambda(p^\mu)$
and $\rho(p^\mu)$ are readily seen to be eigenstates of
the operator
\begin{equation}
{\eta}\,\equiv\,-\,\Gamma^5\,{h}\quad.
\end{equation}
We shall call $ \eta$ a {\it chiral helicity operator}
and its eigenvalues {\it chiral helicities}.

\bigskip
\noindent
{\it Explicit Construction of ${\lambda({ p}^\mu)}$ for Spin-${1/2}$ and
Spin-${1}$ }
\bigskip

The general form of the $(1/2,\,0)\oplus(0,\,1/2)$ and
$(1,\,0)\oplus(0,\,1)$ representation-space $\lambda(p^\mu)$ rest spinors is
obtained from (\ref{os}) by setting $p^\mu\,=\,\overcirc{p}^\mu$ and
equating $\zeta_\lambda$ in accordance with their values determined above.
Once the rest spinors $\lambda(\overcirc{p}^\mu)$ are written down, the
$\lambda(p^\mu)$ follow by the application of an appropriate Wigner boost:
\begin{equation}
\lambda(p^\mu)\,=\,W(j,\, p^\mu\leftarrow\overcirc{p}^\mu)\,
\lambda(\overcirc{p}^\mu)\quad,\label{wb}
\end{equation}
where (Using Eq. (\ref{wjrl}) and appropriate expansions
of $\exp\left(\pm\,{\bf J}\,\cdot\bbox{\varphi}\right)$.)
\footnotemark[8]
\footnotetext[8]
{It is a noteworthy observation that the helicity operator ${\bf J\cdot
\widehat{\bf p}}$ enters the Wigner boosts (\ref{w}) and (\ref{wo}) in a
non-trivial fashion. The linearity of the spin-$1/2$ Dirac equation in
$\partial_\mu$, and the quadratic nature of the  modified Weinberg equation
\cite{SW,FNBWW}, lie in this observation. The specific difference arises as
a result of $j$-dependent behavior of $({\bf J\cdot p})^n$;  $n\,=$
integer.}
\begin{eqnarray}
&&W(1/2,\,p^\mu\leftarrow\overcirc{p}^\mu)\,=\,
\left({{E\,+\,m}\over{2\,m}}\right)^{1/2}\,
\left(
\begin{array}{cc}
\openone\,+\,(E\,+\,m)^{-\,1}\,\bbox{\sigma\cdot p} & 0 \\
0 & \openone\,-\,(E\,+\,m)^{-\,1}\,\bbox{\sigma\cdot p}
\end{array}\label{w}
\right)\quad,\\
&&W(1,\,p^\mu\leftarrow\overcirc{p}^\mu)\,=\, \nonumber\\
&&\,\,\left(
\begin{array}{cc}
\openone\,+\,m^{-1}\,{\bf J\cdot p}\,+\,{\bbox(}m\,(E\,+\,m){\bbox)}^{-1}
\,({\bf J\cdot p})^2 & 0\\
0 & \openone\,-\,m^{-1}\,{\bf J\cdot p}\,+\,{\bbox(}m\,(E\,+\,m){\bbox)}^{-1}
\,({\bf J\cdot p})^2
\end{array}
\right)
\quad.\label{wo}
\end{eqnarray}
To construct $\lambda(\overcirc{p}^\mu)$ that  are eigenstates of the
chiral helicity operator, we choose $\phi_{_L}(\overcirc{p}^\mu)$
to be eigenstates of ${\bf J\cdot \widehat{\bf p}}$.

The explicit expressions for the $\lambda(p^\mu)$ using the above
procedure, and exploiting definition
(\ref{ida}) and identity (\ref{idb}), yields the following results:

 \bigskip \noindent {\it I. For Spin-${1/2}$} \begin{mathletters}
\begin{eqnarray} &&\lambda_\uparrow^S(p^\mu)\,=\,
\left({{E\,+\,m}\over{2\,m}}\right)^{1/2}\, \left( \begin{array}{c}
\Bigl\{1\,-\,(E\,+\,m)^{-\,1}\,|{\bf{p}}|\Bigr\}\,\,i\,\Theta_{[1/2]}\,
\left[\phi^{+\, 1/2}_{_L}(\overcirc{p}^\mu)\right]^\ast\\ \Bigl\{1
\,-\,(E\,+\,m)^{-\,1}\,|{\bf{p}}|\Bigr\}\,\phi^{+\,1/2}
_{_L}(\overcirc{p}^\mu)
\end{array}
\right)\label{lspa}\quad,\\
&&\lambda_\downarrow^S(p^\mu)\,=\,
\left({{E\,+\,m}\over{2\,m}}\right)^{1/2}
\left(
\begin{array}{c}
\Bigl\{1\,+\,(E\,+\,m)^{-\,1}\,|{\bf{p}}|\Bigr\}\,\,i\,\Theta_{[1/2]}\,
\left[\phi^{-\,1/2}_{_L}(\overcirc{p}^\mu)\right]^\ast\\
\Bigl\{1 \,+\,(E\,+\,m)^{-\,1}\,|\bf{p}|\Bigr\}\,\phi^{-\,1/2}
_{_L}(\overcirc{p}^\mu)
\end{array} \right)\quad.\label{lspb} \end{eqnarray}
\end{mathletters}

\bigskip

\noindent
{\it II. For Spin-$ 1$}
\begin{mathletters}
\begin{eqnarray}
&&\lambda_\uparrow^S(p^\mu)\,=\,
\left(\begin{array}{c}
\Bigl\{\openone\,-\,m^{-1}\,\vert {\bf p} \vert \,+\,
{\bbox(}m\,(E\,+\,m){\bbox)}^{-1}\vert {\bf p} \vert^2 \Bigr\}\,
\Theta_{[1]}\, \left[\phi^{+1}_{_L}(\overcirc{p}^\mu)\right]^\ast \\
\Bigl\{\openone\,-\,m^{-1}\,\vert {\bf p} \vert \,+\,
{\bbox(}m\,(E\,+\,m){\bbox)}^{-1}\vert {\bf{p}} \vert^2 \Bigr\}\,
\phi^{+1}_{_L}(\overcirc{p}^\mu)
\end{array}\right) \quad,\label{lsopa}\\
&&\lambda_\rightarrow^S(p^\mu)\,=\,
\left(
\begin{array}{c}
\Theta_{[1]}\, \left[\phi^{0}_{_L}(\overcirc{p}^\mu)\right]^\ast \\
\phi^{0}_{_L}(\overcirc{p}^\mu)
\end{array}
\right)\quad,\label{lsopb}\\
&&\lambda_\downarrow^S(p^\mu)\,=\,
\left(
\begin{array}{c}
\Bigl\{\openone\,+\,m^{-1}\,\vert {\bf p} \vert \,+\,
{\bbox(}m\,(E\,+\,m){\bbox)}^{-1}\vert {\bf{p}} \vert^2 \Bigr\}\,
\Theta_{[1]}\, \left[\phi^{-1}_{_L}(\overcirc{p}^\mu)\right]^\ast \\
\Bigl\{\openone\,+\,m^{-1}\vert {\bf p} \vert \,+\,
{\bbox(}m\,(E\,+\,m){\bbox)}^{-1}\vert {\bf{p}} \vert^2 \Bigr\}\,
\phi^{-1}_{_L}(\overcirc{p}^\mu)
\end{array}\right) \quad. \label{lsopc}
\end{eqnarray}
\end{mathletters}

For spin-$1/2$, the subscripts $\uparrow$ and $\downarrow$ correspond to
$\lambda(\overcirc{p}^\mu)$ constructed out of
$\phi^+_{_{L}}(\overcirc{p}^\mu)$ and $\phi^-_{_{L}}(\overcirc{p}^\mu)$,
respectively, and are to be interpreted  as  chiral helicities. Similarly
for spin-$1$, the subscripts $\uparrow$, $\rightarrow$, and $\downarrow$
correspond to $\lambda(\overcirc{p}^\mu)$ constructed out of
$\phi^{+1}_{_{L}}(\overcirc{p}^\mu)$, $\phi^0_{_{L}}(\overcirc{p}^\mu)$,
and  $\phi^{-1}_{_{L}}(\overcirc{p}^\mu)$, respectively.

The expressions for the anti-self charge-conjugate spinors
$\lambda^A(p^\mu)$ are obtained by replacing $i\,\Theta_{[1/2]}\,$ by
$-\,i\,\Theta_{[1/2]}\,$, for spin-$1/2$, and $\Theta_{[1]}\,$ by
$-\,\Theta_{[1]}\,$, for spin-$1$,  in the above expressions and at the
same time replacing $\lambda^S(p^\mu)$ by $\lambda^A(p^\mu)$ without
changing the chiral helicity indices.

If we restrict ourselves to the physically acceptable
norms, $N^2$, for $\phi^\pm(p^\mu)$ such
that for {\it massless} particles, all {\it rest}-spinors vanish (because
massless
particles cannot be at rest);
\footnotemark[9]
\footnotetext[9]{A convenient choice \cite{FNBWW,MSa}
satisfying this requirement is $N=m^j\,\times\,
(\mbox{phase}\,\,\mbox{factor})$.} then, first considering spin-$1/2$,
an inspection of $\lambda(p^\mu)$
given by Eqs. (\ref{lspa}) and (\ref{lspb}) immediately reveals that for
massless particles  there exists a kinematical asymmetry for the
self/anti-self charge-conjugate spinors in the $(1/2,\,0)\oplus(0,\,1/2)$
representation space. For massless particles, $\lambda_\uparrow^S(p^\mu)$
$\bbox($and $\lambda_\uparrow^A (p^\mu)$$\bbox)$  identically vanish.
However, this vanishing should not be associated with the norm we have
chosen. The norm simply avoids the unphysical and singular norm of the
massless spinors. The physical origin of this asymmetry lies  in the fact
that $\Bigl\{\openone\,+\,(E\,+\,m)^{-\,1}\,\bbox{\sigma\cdot p}\Bigr\}$,
which appears in the Wigner boost (\ref{w}), acting on $\Theta_{[1/2]}\,
\left[\phi^+_{_L}(\overcirc{p}^\mu)\right]^\ast$, a factor that originates
from the requirement of self/anti-self charge conjugacy, on exploiting the
identity (\ref{idb}), which has its origin in the very specific property of
the Wigner's  operator
($
\Theta_{[j]}\,{\bf J}\,\Theta_{[j]}^{-1}\,=\,-\,{\bf J}^\ast
$), conspire to yield
$\Bigl\{\openone \,-\,(E\,+\,m)^{-\,1}\,|{\bf{p}}|\Bigr\}\,\,\Theta_{[1/2]}\,
\left[\phi^+_{_L}(\overcirc{p}^\mu)\right]^\ast$, which for the massless
case vanishes. Similar remarks also apply to spin-$1$.

The most general forms for the $\phi^{h^\prime}(\overcirc{p}^\mu)$ that
appear in expressions (\ref{lspa}) to (\ref{lsopc}) are:

\bigskip

\noindent
{\it A. For Spin-${1/2}$}
\begin{mathletters}
\begin{eqnarray}
&&\phi^{+\,1/2}_{_L}({\overcirc{p}}^\mu)\,=\,N\,e^{i\,\vartheta_1}\,
\left(\begin{array}{l}
\cos(\theta/2)\,e^{-i\,\phi/2}\\
\sin(\theta/2)\,e^{i\,\phi/2}
\end{array}\right)
\quad,\label{exphia}\\
&&\phi^{-\,1/2}_{_L}({\overcirc{p}}^\mu)\,=\,N\,e^{i\,\vartheta_2}\,
\left(\begin{array}{c}
\sin(\theta/2)\,e^{-i\,\phi/2}\\
-\,\cos(\theta/2)\,e^{i\,\phi/2}
\end{array}\right)\quad.\label{exphib}
\end{eqnarray}
\end{mathletters}

\bigskip

\noindent
{\it B. For Spin-${1}$}
\begin{mathletters}
\begin{eqnarray}
&&\phi^{+\,1}_{_L}({\overcirc{p}}^\mu)\,=\,{\rm N}\,e^{i\,\delta_1}\,
\left(
\begin{array}{l}
{1\over 2}\,\left(1\,+\,\cos\,\theta\right)\,e^{-i\,\phi}\\
\sqrt{1\over 2}\,\sin\,\theta\\
{1\over 2}\,\left(1\,-\,\cos\,\theta\right)\,e^{i\,\phi}
\end{array}\right)\quad,\\
&&\phi^{0}_{_L}({\overcirc{p}}^\mu)\,=\,{\rm N}\,e^{i\,\delta_2}\,
\left(\begin{array}{l}
-\,\sqrt{1\over 2}\,\sin\,\theta\,e^{-i\,\phi}\\
\cos\,\theta\\
\sqrt{1\over 2}\,\sin\,\theta\,e^{i\,\phi}
\end{array}\right)\quad,\\
&&\phi^{-\,1}_{_L}({\overcirc{p}}^\mu)\,=\,{\rm N}\,e^{i\,\delta_3}\,
\left(\begin{array}{l}
{1\over 2}\,\left(1\,-\,\cos\,\theta\right)\,e^{-i\,\phi}\\
-\,\sqrt{1\over 2}\,\sin\,\theta\\
{1\over 2}\,\left(1\,+\,\cos\,\theta\right)\,e^{i\,\phi}
\end{array}\right)\quad.
\end{eqnarray}
\end{mathletters}
Here, $\theta$ and $\phi$ are the
standard polar and azimuthal angles associated with $\bf{p}$.

Since both $\phi_{_L}(\overcirc{p}^\mu)$ and its complex conjugate
$\phi^\ast_{_L}(\overcirc{p}^\mu)$ enter a given $\lambda(p^\mu)$,
 the covariant norm $\overline{\lambda}(p^\mu)\,\lambda(p^\mu)$
depends on the choice of phase factors $e^{\vartheta_{1,2}}$ and
$e^{\delta_{1,2,3}} $.
\footnotemark[10]
\footnotetext[10]{Recall that
$\overline{\psi}(p^\mu)$ in the $(j,\,0)\oplus(0,\,j)$ representation space
is defined as $\psi^\dagger(p^\mu)\,\times \left(\begin{array}{cc} 0
&\openone\\ \openone &  0\end{array}\right)$; where
$\openone\,=\,\mbox{a}\,\,(2\,j\,+\,1)\times(2\,j\,+\,1)$ identity matrix.}
This is
manifest in  Tables I and II, where
$\overline{\lambda}(p^\mu)\,\lambda(p^\mu)$ are explicitly tabulated. Any
phase choice for which $\vartheta_1\,+\,\vartheta_2$ vanishes yields a
bi-orthonormal set for spin-$1/2$. Similarly, for spin-$1$, phase choices
for which $\delta_3\,+\,\delta_1$ and $\delta_2$ vanish yield
a bi-orthonormal set. We shall make these choices in the rest of this paper.
In fact, for convenience, we set $\vartheta_1\,= 0\,=\,\vartheta_2$ for
spin-$1/2$; and $\delta_1\,=\,\delta_2\,=\,0\,=\delta_3$ for spin-$1$.
For spin-$1$, as well as for spin-$1/2$, $\lambda^S(p^\mu)$
and $\lambda^A(p^\mu)$
span mutually orthogonal sub-spaces, with the completeness relations given
by:
\begin{eqnarray}
{-1\over{2\,i\,N^2}}{\biggl[}{\Bigl\{}
\lambda^S_\uparrow(p^\mu)\,\overline{\lambda}^S_\downarrow(p^\mu)&\,-\,&
\lambda^S_\downarrow(p^\mu)\,\overline{\lambda}^S_\uparrow(p^\mu)
{\Bigr\}}
\,-\,\nonumber\\
&&{\Bigl\{}
\lambda^A_\uparrow(p^\mu)\,\overline{\lambda}^A_\downarrow(p^\mu)\,-\,
\lambda^A_\downarrow(p^\mu)\,\overline{\lambda}^A_\uparrow(p^\mu)
{\Bigr\}} {\biggr]}\,=\,\openone\quad,
\end{eqnarray}
for spin-$1/2$; and for spin-$1$ by
\begin{eqnarray}
{1\over {2\,{\rm N}^2}}
{\biggl[}{\Bigl\{}\lambda^S_\uparrow(p^\mu)\,\overline{\lambda}^S_
\downarrow(p^\mu)&\,-\,&
\lambda^S_\rightarrow(p^\mu)\,\overline{\lambda}^S_\rightarrow(p^\mu) \,+\,
\lambda^S_\downarrow(p^\mu)\,\overline{\lambda}^S_\uparrow(p^\mu) {\Bigr\}}
\,-\,\nonumber \\
&&{\Bigl\{}\lambda^A_\uparrow(p^\mu)\,\overline{\lambda}^A_
\downarrow(p^\mu)\,-\,
\lambda^A_\rightarrow(p^\mu)\,\overline{\lambda}^A_\rightarrow(p^\mu) \,+\,
\lambda^A_\downarrow(p^\mu)\,\overline{\lambda}^A_\uparrow(p^\mu) {\Bigr\}}
{\biggr]}\,=\,\openone\quad.
\end{eqnarray}

The construction of the $\rho(p^\mu)$ spinors follows in a parallel fashion.

\bigskip
\noindent{\it New Wave Equations for Spin-$1/2$ and Spin-$1$:
 Wave Equation for ${\lambda({p}^\mu)}$}

To obtain a wave equation satisfied by $\lambda(p^\mu)$ we must first
generalize the Ryder-Burgard relation. This is accomplished be defining a
$(2j+1)\times(2j+1)$ matrix ${\mit\Xi}_{[j]}$ such that
\footnotemark[11]
\footnotetext[11]{ {\it Cf.} Equation (\ref{br}) with
the equation in the second line on p. 44 of Ryder's book \cite{LHR}. The
corrected form of the indicated equation of Ryder was obtained in
discussions with Burgard \cite{CB}
and can be found in Ref. \cite{FNBWW}. In
part, the analysis of Ref. \cite{FNBWW} yields  {\it Dirac}-like  modified
Weinberg wave equations for $(j,\,0)\oplus(0,\,j)$ spinors.}
\begin{equation}
\left[\phi^{h^\prime}_{_{L}}({\overcirc{p}}^\mu)\right]^\ast\,=\,
{\mit\Xi}_{[j]}\,\phi^{h^\prime}_{_{L}}({\overcirc{p}}^\mu)\quad.\label{br}
\end{equation}
With the choice of phases $e^{i\vartheta}$ and $e^{i\delta}$ introduced
above, ${\mit\Xi}_{[j]}$ for spin-$1/2$ and spin-$1$ read:
\begin{equation}
{\mit\Xi}_{[1/2]}\,=\,\left(\begin{array}{cc}
e^{i\phi} & 0\\
0 & e^{-i\phi}
\end{array}\right)\quad,\qquad
{\mit\Xi}_{[1]}\,=\,\left(\begin{array}{ccc}
e^{i\,2\,\phi}&0 & 0\\
0&1&0\\
0 & 0 & e^{-i\,2\,\phi}
\end{array}\right)\quad.
\end{equation}
Next, for convenience, we introduce the abbreviation
\begin{equation}
\chi_{_{R}}(p^\mu)\,\equiv\,
\left(\zeta_\lambda\,\Theta_{[j]}\right)\,\phi^\ast_{_L}(p^\mu)
\label{chir}
\end{equation}
and couple the transformation properties
\begin{mathletters}
\begin{eqnarray}
\chi_{_{R}}(p^\mu) & \,=\,&
\exp\left(\,{\bf{J}}\cdot\bbox{\varphi}\right)
\,\chi_{_{R}}({\overcirc{p}}^\mu)\quad,\\
\phi_{_L}(p^\mu) & \,=\,&
\exp\left(-\,{\bf{J}}\cdot\bbox{\varphi}\right)
\,\phi_{_L}({\overcirc{p}}^\mu)\quad,
\end{eqnarray}
\end{mathletters}
\noindent
with (\ref{br}) to obtain (after a short string of algebraic manipulations
that exploit some already indicated  properties of $\Theta_{[j]}$ and
$\zeta_\lambda$ and are similar in nature to those found in Sec. 2.3 of
Ryder's book \cite{LHR} and Refs. \cite{FNBWW,Vorticity})
\begin{equation}
\left(
\begin{array}{cc}
-\,\openone & \zeta_\lambda\,\exp\left( {\bf J}\,\cdot\bbox{\varphi}\right)
\,\Theta_{[j]}\,{\mit\Xi}_{[j]}\, \exp\left( {\bf J}\,\cdot\bbox{\varphi}
\right)\\
\zeta_\lambda\,\exp\left(-\, {\bf J}\,\cdot\bbox{\varphi}\right)
\,{\mit\Xi}^{-1}_{[j]}\,\Theta_{[j]}\, \exp\left(- \,{\bf
J}\,\cdot\bbox{\varphi}
\right) & -\,\openone \end{array}
\right)\,\lambda(p^\mu)\,=\,0\quad.\label{genweq} \end{equation} This is
the general wave equation satisfied by  arbitrary-spin $\lambda(p^\mu)$.

For spin-$1/2$ the equation (\ref{genweq}) takes the form
\begin{equation}
\left(
\begin{array}{cc}
-\,2\,m\,(E\,+\,m)\,\openone  &
\zeta_\lambda\,\left\{ (E\,+\,m)\,\openone \,+\,
\,\bbox{\sigma\cdot}{\bf{p}}\right\}
\Theta_{[1/2]}\,{\mit\Xi}_{[1/2]} \\
{}& \times\,\left\{ (E\,+\,m)\,\openone \,+\,
\,\bbox{\sigma\cdot}{\bf{p}}\right\}\\
\zeta_\lambda\,\left\{ (E\,+\,m)\,\openone \,-\,
\,\bbox{\sigma\cdot}{\bf{p}}\right\}\,{\mit\Xi}^{-1}_{[1/2]}\,\Theta_{[1/2]}
& {} \\ \times\,\left\{ (E\,+\,m)\,\openone \,-\,
\,\bbox{\sigma\cdot}{\bf{p}}\right\}
& -\,2\,m\,(E\,+\,m)\,\openone
\end{array}\right)\,\lambda(p^\mu)\,=\,0\quad;\label{weqh}
\end{equation}
and for spin-$1$ it becomes \begin{equation}
\left( \begin{array}{cc} -\,\left\{m\,(m\,+\,E)\right\}^2\,\openone &
\zeta_\lambda\,\{\cdots\,A_+\,\cdots\} \,\Theta_{[1]}\,{\mit\Xi}_{[1]} \,
   \{\cdots\,A_+\,\cdots\}      \\ \zeta_\lambda\, \{\cdots\,A_-\,\cdots\}
{\mit\Xi}^{-1}_{[1]}\,\Theta_{[1]} \{\cdots\,A_-\,\cdots\} &
-\,\left\{m\,(m\,+\,E)\right\}^2 \,\openone
\end{array}\right)\,\lambda(p^\mu)\,=\,0\quad,\label{weqo}
\end{equation}
with
\begin{equation}
\{\cdots\,A_\pm\,\cdots\}\,=\,
\left\{\left(\eta_{\mu\nu}p^\mu p^\nu\,+\,m\,E\right)\,\openone
\,\pm\,(m\,+\,E)\,{\bf J\cdot p}\,+\,\left({\bf J\cdot p}\right)^2\right\}
\quad,
\end{equation}
and $\eta_{\mu\nu}$ is the flat space-time metric with
${\mbox{diag}}(1,\,-1,\,-1,\,-1)$.

Both for spin-$1/2$ and spin-$1$, using the symbolic manipulation program
DOEMACSYMA, we find that while $\lambda^S(p^\mu)$ are the {\it positive}
energy solutions with $E\,=\,+\,\sqrt{m^2\,+\,{\bf{p}}^2}\,$,
$\lambda^A(p^\mu)$ are the {\it negative} energy solutions with
$E\,=\,-\,\sqrt{m^2\,+\,{\bf{p}}^2}\,$. These results suggest a
particle-antiparticle interpretation. Since the particle-hole picture so
useful for fermions cannot be generalized to bosons (because filling the
negative energy sea works for fermions but not for bosons), we shall,
following Hatfield \cite{BH}, interpret negative energy particles (fermions
or bosons) propagating backward in time as antiparticles in accordance with
the St\"ukelberg/Feynman-Wheeler \cite{SF} picture of space-time. Similar
results are obtained for $\rho(p^\mu)$.

For type II spinors, it should be explicitly noted that, unlike the case
for type I spinors, the Charge Conjugation operator does not take the
positive (negative) energy solutions into negative (positive) energy
solutions.

Before proceeding further, we make a few
brief remarks on  the existence of more than one equation (unconnected
by  unitary transformations) in a given representation space. As early as
1932 Majorana \cite{EMa} obtained a relativistic wave equation for
spin-$1/2$ particles (which included a tower of higher spins also) starting
from the demand that {\it all} its solutions have positive energy. Almost
four decades later in 1971 Dirac \cite{PAMDa} proposed a relativistic wave
equation for bosons, which again allowed only positive-energy solutions {\it
and} had  the property that it described particles  that did not
interact electromagnetically (see Dirac's comments on pp. 68-69 in Ref.
\cite{PAMDb}). The reason that we obtain an additional wave equation in the
$(1/2,\,0)\oplus(0,\,1/2)$  representation space that is different from
Dirac's famous equation of 1928 {\it and} a new wave equation in the
$(1\,,0)\oplus(0,\,1)$ representation space that is different from (the
modified \cite{FNBWW}) Weinberg's wave equation \cite{SW,HJoos} of 1964 is
that wave equations are determined by the (physically-motivated)
constraints that we impose on its solutions and transformation properties
of the objects in the representation space under consideration. In general,
different constraints yield different wave equations. Dirac (1971,1972) and
Majorana (1932) imposed the constraint of positive-energy solutions. For
the solutions we consider, we imposed the requirement of the self/anti-self
charge conjugacy for spin-$1/2$ and self/anti-self conjugacy under the
operation of $\Gamma^5\,S^c_{[1]}$ for spin-$1$. Dirac's famous
equation of 1928 describes  eigenspinors of the Charge operator.

\bigskip \noindent
{\it Some Considerations on Parity}

We begin with the  study of parity covariance of
equations (\ref{weqh}) and (\ref{weqo}). Define the operator that acts on
$\lambda(p^\mu)$ in the equation (\ref{weqh}) as ${\cal
O}_{[1/2]}\,(p^\mu)$. Similarly, referring to (\ref{weqo}), introduce
${\cal O}_{[1]}\,(p^\mu)$. The parity covariance of equations ${\cal
O}_{[1/2]}\,(p^\mu)\,\lambda(p^\mu)\,=\,0$ and ${\cal
O}_{[1]}\,(p^\mu)\,\lambda(p^\mu)\,=\,0$ demands that we seek operators
${\bbox{\cal S}}^s_{[\alpha]}$ such that under $\cal P$
\begin{eqnarray} {\cal O}_{[\alpha]}(p^\mu) \,&&\rightarrow\,{\cal
O}_{[\alpha]}^\prime(p^\mu)\,=\,{\bbox{\cal S}}^s_{[\alpha]}\,{\cal
O}_{[\alpha]}(p^\mu)\,\left[ {\bbox {\cal S}}^s_{[\alpha]}\right]^{\,-1}
\quad, \\
\lambda(p^\mu)\,&&\rightarrow\,\lambda'(p^{\mu})\,=\,
{\bbox{\cal S}}^s_{[\alpha]}\,\lambda(p^\mu)
\quad \alpha\,=\,1/2,\,1\quad;
\end{eqnarray}
with
${\cal
O}_{[\alpha]}^\prime(p^\mu)\,=\,{\cal O}_{[\alpha]} (p^{\prime\,\mu})$
and $\lambda'(p^{\mu})\,=\,\lambda(p^{\prime\,\mu})\,$.
Here,
$p^{\prime\,\mu}$ is the parity-transformed $p^\mu$
and reads $(E,\,-\,{\bf p})$ for $p^\mu\,=\,(E,\,{\bf p})$. On exploiting
the fact that $
\Theta_{[\alpha]}\,{\mit\Xi}_{[\alpha]}\,=\,
{\mit\Xi}_{[\alpha]}^{-\,1}\,\Theta_{[\alpha]}\,, $ such  operators
are found to be identical to those already given in (\ref{cph}) and
(\ref{cpo}); that is: ${\bbox{\cal S}}^s_{[\alpha]}\,=\,
S^s_{[\alpha]}$.
This is not surprising. Just as the operator of parity in the
$(j,\,0)\oplus(0,\,j)$ representation space is independent of which wave
equation is under study, similarly the operations of charge conjugation and
time reversal
\footnotemark[12]
\footnotetext[12] { The charge conjugation
and parity operations in the $(j,\,0)\oplus(0,\,j)$ representation space
for spin-$1/2$ and spin-$1$ are given by (\ref{cph}) and (\ref{cpo}). These
are supplemented by the time-reversal operators
\cite{ON,FNBWW} (within a
global phase factor)
\[
S_{[1/2]}(T)\,=\,-\,\left(\begin{array}{cc}
0&\Theta_{[1/2]}\\
\Theta_{[1/2]}&0
\end{array}\right)\quad,\qquad
S_{[1]}(T)\,=\,\,\left(\begin{array}{cc}
\Theta_{[1]} & 0\\
0&-\,\Theta_{[1]}
\end{array}\right)\quad.
\]
However, it should be again recalled that the analysis by Nigam
and Foldy reveals that the Parity operator in the Fock space depends on the
Charge operator.}
 do not depend on a
specific wave equation.
Within the context of the logical framework of the
present paper, without this being true we would not even
know  how to define
self-/anti self conjugate $(j,\,0)\oplus(0,\,j)$ spinors.
These remarks should not be interpreted to mean that there may not arise
certain subtle differences between the operations of P and C
in the Fock space \cite{NF}.

We no longer pursue the subject of CPT covariance any further. A detailed
analysis of these operators, within the context of  a
Foldy-Nigam-Bargmann-Wightman-Wigner type quantum field theory, was
recently published in Ref. \cite{FNBWW}. To incorporate the Nigam and Foldy
considerations in the work of Ref. \cite{FNBWW} one simply follows the
details of \cite{NF} in a straightforward manner.

With the choice of phases $e^{i\,\vartheta}$ and $e^{i\,\delta}$ made
above, the effect of the parity operator, $S^s_{[j]}$, on
$\lambda^S(p^\mu)$ and $\lambda^A(p^\mu)$ spinors is given as follows:

\bigskip

\noindent
{\it ${{\alpha}}$. For Spin-${1/2}$}

\begin{mathletters}
\begin{eqnarray}
&&\gamma^0\,\lambda^S_\uparrow(p^\mu)\,=\,-\,i\,\lambda^S_\downarrow
(p^{\prime\,\mu})\,\,, \quad
\gamma^0\,\lambda^S_\downarrow(p^\mu)\,=\,+\,i\,\lambda^S_\uparrow
(p^{\prime\,\mu})\quad,\\
&&\gamma^0\,\lambda^A_\uparrow(p^\mu)\,=\,+\,i\,\lambda^A_\downarrow
(p^{\prime\,\mu})\,\,, \quad
\gamma^0\,\lambda^A_\downarrow(p^\mu)\,=\,-\,i\,\lambda^A_\uparrow
(p^{\prime\,\mu})\quad,\label{gol}
\end{eqnarray}
\end{mathletters}

\noindent
{\it ${{\beta}}$. For Spin-${1}$}

\begin{mathletters}
\begin{eqnarray}
&&\gamma_{00}\,\lambda^S_\uparrow(p^\mu)\,=\,+\,\lambda^S_\downarrow
(p^{\prime\,\mu})\,\,, \quad
\gamma_{00}\,\lambda^S_\rightarrow(p^\mu)\,=\,-\,\lambda^S_\rightarrow
(p^{\prime\,\mu})\,\,, \quad
\gamma_{00}\,\lambda^S_\downarrow(p^\mu)\,=\,+\,\lambda^S_\uparrow
(p^{\prime\,\mu})\quad, \\
&&\gamma_{00}\,\lambda^A_\uparrow(p^\mu)\,=\,-\,\lambda^A_\downarrow
(p^{\prime\,\mu})\,\,, \quad
\gamma_{00}\,\lambda^A_\rightarrow(p^\mu)\,=\,+\,\lambda^A_\rightarrow
(p^{\prime\,\mu})\,\,, \quad
\gamma_{00}\,\lambda^A_\downarrow(p^\mu)\,=\,-\,\lambda^A_\uparrow
(p^{\prime\,\mu})\quad .
\end{eqnarray}
\end{mathletters}

The $\lambda(p^\mu)$ for spin-$1/2$ are not eigenspinors of the parity
operator. This is not related to the fact that $S_{[1/2]}({\cal C})$ and
$S_{[1/2]}({\cal P})$ do not commute. Since $S_{[1/2]}({\cal C})$ is not
linear, it is possible to have a simultaneous set of eigenspinors for
$S_{[1/2]}({\cal C})$ and $S_{[1/2]}({\cal P})$; but such a set does not
have its eigenspinors of type II.
\footnotemark[13]
\footnotetext[13]
{This argument was constructed with Christoph Burgard and George
Kahrimanis. I thank them both.}

The $\lambda(p^\mu)$ for spin-$1$ can however be made into
eigenspinors of the Parity operator ({\it without} destroying the
$\theta$-conjugacy). Specifically,
\begin{equation}
\lambda^S_\pm(p^\mu)\,\equiv{1\over\sqrt{2}}\biggl[\lambda_\uparrow^S(p^\mu)\,\pm\,
\lambda_\downarrow^S(p^\mu)\biggr]\,\,,\quad\lambda_0(p^\mu)\,\equiv\,
\lambda_\rightarrow^S(p^\mu)\quad,
\end{equation}
[with a similar expressions for the $\lambda_{\pm,0}^A(p^\mu)\,$] are
simultaneously eigenspinors of the $\Gamma^5\,S_{[1]}({\cal C})$ and
$S_{[1]}({\cal P})$ operators:
\begin{eqnarray}
&&\Gamma^5\,S_{[1]}({\cal C})\,\lambda^S_{\pm,0}(p^\mu)\,=\,
\lambda^S_{\pm,0}(p^\mu)\quad,\\ &&\gamma_{00}\,\lambda^S_\pm(p^\mu)\,=\,
+\,\lambda^S_\pm(p^{\prime\mu})\,\,,\quad
\gamma_{00}\,\lambda^S_0(p^\mu)\,=\, -\,\lambda^S_0(p^{\prime\mu})\quad.
\end{eqnarray}
The subscript $\pm$ no longer
corresponds to the chiral helicity.
$\lambda^S_\pm(p^\mu)$ and $\lambda^S_0(p^\mu)$ carry opposite relative
intrinsic parities [with similar statement being true for
$\lambda_{\pm,0}^A(p^\mu)\,$]. If $\lambda_{\pm,0}(p^\mu)$ were realized as
physical degrees of freedom then any interaction that induces transitions
between the $\pm$ and $0$ degree of freedom would necessarily violate
parity, except in the massless case where transitions to the $0$ degree
of freedom would be expected to vanish identically on the basis of
arguments presented in the section entitled
``Explicit Construction of ${\lambda({ p}^\mu)}$ for Spin-${1/2}$ and
Spin-${1}\,$.''

\bigskip
\noindent
{\it Dirac-Like and Majorana-Like Fields}

On setting
$\vartheta_1\,= 0\,=\,\vartheta_2$ for
spin-$1/2$, and $\delta_1\,=\,\delta_2\,=\,0\,=\delta_3$ for spin-$1$,
as discussed before,
the $\lambda(p^\mu)$ form a bi-orthogonal set of self/anti-self
charge conjugate spinors. Similar results hold true for $\rho^S(p^\mu)$ and
$\rho^A(p^\mu)$. In addition, there exist several identities that  relate
$\lambda^S(p^\mu)$ and $\rho^A(p^\mu)$, etc. (see, e.g., Eqs. (48a)
and (48b) below).

Since the set of self/anti-self charge conjugate spinors cannot be made
orthonormal without destroying self/anti-self charge conjugacy, the noted
bi-orthonormality suggests that the physical states be normed as:
\begin{equation}
^{\beta^\prime}\langle\, p^{\prime\,\mu},\,\eta^\prime\,\vert \,
 p^\mu,\,\eta \,\rangle^{\beta}\,=\,
(2\,\pi)^3 \, 2\,p_0\,\delta^3({\bf p}\,-\,{\bf p}^\prime)\,
\delta_{\eta,\,-\,\eta^\prime}\,\delta_{\beta\,\beta^{\prime}}\quad.
\label{states}
\end{equation}
Notational Note: $\pm\,\uparrow\,=\,\mp\,\downarrow$ and
$-\,\rightarrow\,=\,\rightarrow\,$; $\beta$ and $\beta^\prime$
can take values $S$ and $A$, which refer to the self and anti-self
charge conjugacy identifying indices.
Consider $\beta\,=\,\beta^\prime\,=\,S$, first.
The {\it lhs} of (\ref{states}) can now be written as
\begin{equation}
\left\langle\mbox{vac}\left\vert\,\left[a_{\eta^\prime}( p^{\prime\,\mu}),\,
a^\dagger_\eta(p^\mu)\right]_\pm\,\right\vert\mbox{vac}\right\rangle \,\mp\,
\left\langle\mbox{vac}\left\vert a^\dagger_\eta(p^\mu)\, a_{\eta^\prime}(
p^{\prime\,\mu})\right\vert\mbox{vac}\right\rangle \,=\, (2\,\pi)^3 \,
2\,p_0\,\delta^3({\bf p}\,-\,{\bf p}^\prime)\,
\delta_{\eta,\,-\,\eta^\prime}\quad,
\end{equation}
where $\vert\mbox{vac}\rangle$ represents the vacuum state.
The second term in the {\it lhs} of the above equation vanishes identically,
yielding
\begin{equation}
\left[a_{\eta^\prime}( p^{\prime\,\mu}),\,
a^\dagger_\eta(p^\mu)\right]_\pm\,=\,(2\,\pi)^3 \,
2\,p_0\,\delta^3({\bf p}\,-\,{\bf p}^\prime)\,
\delta_{\eta,\,-\,\eta^\prime}\quad.
\end{equation}
Next we consider $\beta\,=\,\beta^\prime\,=\,A$, and obtain
\begin{equation}
\left[b_{\eta^\prime}( p^{\prime\,\mu}),\,
b^\dagger_\eta(p^\mu)\right]_\pm\,=\,(2\,\pi)^3 \,
2\,p_0\,\delta^3({\bf p}\,-\,{\bf p}^\prime)\,
\delta_{\eta,\,-\,\eta^\prime}\quad.
\end{equation}
Finally, by considering
$\beta\,=\,S$ and $\beta^\prime\,=\,A$ (
or, $\beta\,=\,A$ and $\beta^\prime\,=\,S$), we obtain
\begin{equation}
\left[a_{\eta^\prime}( p^{\prime\,\mu}),\,
b^\dagger_\eta(p^\mu)\right]_\pm\,=\,0\quad.\label{extra}
\end{equation}
Using any set of two of the $\lambda^S(p^\mu)$, $\lambda^A(p^\mu)$,
$\rho^S(p^\mu)$, and $\rho^A(p^\mu)$ that forms a complete set we can now
introduce the quantum field motivated by reasons surrounding
Eq. (\ref{qf}). Two examples follow (in the phenomenological context one
must remain open to other inherent possibilities in the formalism):
\begin{enumerate}
\item[1.] A Dirac-like field
\begin{equation}
{\nu}^{DL}(x)\,\equiv\,
\int{ {d^3{\bbox{p}}} \over {(2\,\pi)^3}  } {1\over {2 \,p_0}}
\sum_{\eta}
\left[
\lambda^S_{\eta}(p^\mu)\,a_{\eta}(p^\mu)\,\exp(-\,i\,p\cdot  x)
\,+\,
\lambda^A_{\eta}(p^\mu)\,b^\dagger_{\eta}(p^\mu)\,
\exp(+\,i\,p\cdot x)\right]
\quad,\label{psid}
\end{equation}
or, on identifying $b^\dagger_{\eta}(p^\mu)$
with $a^\dagger_{\eta}(p^\mu)$ (in analogy with obtaining the
Majorana field from the Dirac field)
\item[2.] A Majorana-like field
\begin{equation}
{\nu}^{ML}(x)\,\equiv\,
\int{ {d^3{\bbox{p}}} \over {(2\,\pi)^3}  } {1\over {2 \,p_0}}
\sum_{\eta}
\left[
\lambda^S_{\eta}(p^\mu)\,a_{\eta}(p^\mu)\,\exp(-\,i\,p\cdot  x)
\,+\,
\lambda^A_{\eta}(p^\mu)\,a^\dagger_{\eta}(p^\mu)\,
\exp(+\,i\,p\cdot x)\right]
\quad.
\end{equation}
\end{enumerate}
Note that neither ${\nu}^{DL}(x)$ nor ${\nu}^{ML}(x)$ is a self/anti-self
$\theta$-conjugate field. However, both fields describe self/anti-self
$\theta$-conjugate
states and use of one or the other field leads to a
set of phenomenon that does not have a complete overlap.

In reference to the parenthetic remark  bracketed between
Eqs. (\ref{extra})  and (\ref{psid}), we note
the identities (for spin-$1/2$, similar identities exist for spin-$1$)
\begin{mathletters}
\begin{eqnarray}
&&\rho^S_\uparrow(p^\mu)\,=\,-\,i\,\lambda^A_\downarrow(p^\mu)\,\,,\quad
\rho^S_\downarrow(p^\mu)\,=\,+\,i\,\lambda^A_\uparrow(p^\mu)\quad,
\label{extraa}\\
&&\rho^A_\uparrow(p^\mu)\,=\,+\,i\,\lambda^S_\downarrow(p^\mu)\,\,,\quad
\rho^A_\downarrow(p^\mu)\,=\,-\,i\,\lambda^S_\uparrow(p^\mu)\quad,
\label{extrab}
\end{eqnarray}
\end{mathletters}
These identities may be used to incorporate the left- and right-handed
chiral helicities
in the same spin-$1/2$ field. For instance, for this example case, we
re-write ${\nu}^{DL}(x)$ as
\begin{equation}
{\nu}_2^{DL}(x)\,\equiv\,
\int{ {d^3{\bbox{p}}} \over {(2\,\pi)^3}  } {1\over {2 \,p_0}}
\sum_{\eta}
\left[
\lambda^S_{\eta}(p^\mu)\,c_{\eta}(p^\mu)\,\exp(-\,i\,p\cdot  x)
\,+\,
\rho^S_{\eta}(p^\mu)\,d^\dagger_{\eta}(p^\mu)\,
\exp(+\,i\,p\cdot x)\right]
\quad.\label{psidb}
\end{equation}
The $c_{\eta}(p^\mu)$ and $d^\dagger_{\eta}(p^\mu)$ differ from
$a_{\eta}(p^\mu)$ and $b^\dagger_{\eta}(p^\mu)$ with appropriate phase
factors dictated by
Eqs. (\ref{extraa}) and (\ref{extrab}).

As shown already, in the massless limit, since $\lambda_\downarrow(p^\mu)$ and
$\rho_\uparrow(p^\mu)$ are the only surviving spinors the
massless
spin-$1/2$ field ${\nu}_2^{DL}(x)$ contains only the following states:
self-charge conjugate states $\vert p^\mu,\,\downarrow\rangle^S$ of the
left-handed chiral helicity and anti-self charge conjugate states
$\vert p^\mu,\,\uparrow\rangle^A$ of the right-handed chiral helicity.

\bigskip \noindent
{\it Some Considerations on Interactions}

{}From  a formal point of view, the wave equation (\ref{weqh}) may be put in
the form $\left(\Gamma^{\mu\nu}\,p_\mu \,p_\nu\,+\,m\,\Gamma^\mu\, p_\mu
\,-\,2\, m^2\,\openone\right)\,\lambda(p^\mu)\,=0\,. $ However, it turns
out that $\Gamma^{\mu\nu}$ and $\Gamma^\mu$ do not transform as Poincar\'e
tensors. Therefore, the operator that acts on $\lambda(p^\mu)$ in Eq.
(\ref{weqh}) carries only the  indices of the $(1/2,\,0)\oplus(0,\,1/2)$
representation space {\it without} the additional structure, which contains
contraction[s] of a Poincar\'e tensor[s] with an energy momentum four
vector[s] $p_\mu$. This has the consequence that gauge interactions cannot
be introduced by replacing the $\partial_\mu$ by an appropriate
gauge-covariant derivative. To understand this result better, let us see
what happens to the spinors of type I and type II under simple phase
transformation. First, type I spinors: Under the simultaneous
transformations $\phi_{_R}({p}^\mu)
\,\rightarrow\,e^{i\,\alpha(x)}\,\phi_{_R}({p}^\mu)$ and
$\phi_{_L}({p}^\mu) \,\rightarrow\,e^{i\,\alpha(x)}\,\phi_{_L}({p}^\mu)$ on
$\phi_{_R}({p}^\mu)$ and $\phi_{_L}({p}^\mu)$, the spinors pick up an
overall phase factor $e^{i\,\alpha(x)}\,$. The demand for covariance of the
associated equation of motion (i.e., Dirac equation) under this phase
transformation immediately introduces a local $U(1)$ gauge interaction.
Next,   type II spinors (say the $\lambda(p^\mu)$ spinors): Now let
$\phi_{_L}({p}^\mu) \,\rightarrow\,e^{i\,\alpha(x)}\,\phi_{_L}({p}^\mu)$,
then $\chi_{_{R}}(p^\mu)$, introduced in (\ref{chir}) and which enters the
definition of $\lambda(p^\mu)$, transforms into
$e^{-\,i\,\alpha(x)}\,\chi_{_{R}}(p^\mu)$. The transformed
$\lambda(p^\mu)$, as is readily seen, is no longer a self/anti-self charge
conjugate spinor. These arguments do not rule out the existence of gauge
interactions with the particles under consideration, but point to the fact
that the gauge  interactions (if present) will enter the formalism in a
somewhat different fashion.

When the above remarks are considered  for spin one half the obvious
question arises: What relevance does the self/anti-self charge conjugate
$(1/2,\,0)\oplus(0,\,1/2)$ representation space have to neutrinos. How can
particles thus described enter gauge interactions? The answer is that
within the standard framework the neutrino/anti-neutrino does not belong to
the Dirac's $(1/2,\,0)\oplus(0,\,1/2)$ representation space, or the
self/anti-self charge conjugate representation space considered here.
Neutrino/anti-neutrino are described by the  $(\openone\,\pm\,\gamma^5)$
projectors that project out the left-/right- handed Weyl neutrinos [i.e.
$(0,\,1/2)$ or $(1/2,\,0)$ representation spaces]. So unless there exist
interactions beyond the standard framework without the
$(\openone\,\pm\,\gamma^5)$ projectors involved one cannot distinguish
between Diracness (either as a Dirac field or a Majorana field) or the
self/anti-self charge conjugateness. If the neutrino participates in
interactions beyond the standard model then alone does the construct
presented here become relevant. In such a scenario the most natural
interaction that arises is of the  type ${\overline{\nu}}(x)\,
(\cos\theta\,+\,i\,\gamma^5\sin\theta) \nu(x)\,\phi(x)$, where $\phi(x)$ is
some scalar field (perhaps of the type suggested by Stephenson and Goldman
\cite{SG}) and $\theta$ determines the behaviour of the interaction under
combined operation of charge conjugation and parity.

\bigskip
\noindent
{\it Concluding Remarks}
\bigskip

In view of the considerations presented above we conclude that  the subject
of $(1,\,3)$ space-time symmetries and its implications for the kinematic
structure of quantum field theories is still an open arena. As we have
already pointed out, we have not investigated all possible physically
relevant requirements that may be used to fix relative phase factors
between the $(j,\,0)$ and $(0,\,j)$ spinors. For instance, we may begin with
spinors \begin{equation} \psi(p^\mu)\,=\, \left(\begin{array}{c}
\phi_{_R}(p^\mu)\\ \cos(\theta)\,\phi_{_L}(p^\mu)
\end{array}\right)\,\,,\,\, \mbox{or}\quad \psi(p^\mu)\,=\,
\left(\begin{array}{c}
\cos(\theta)\,\phi_{_R}(p^\mu)\\
\phi_{_L}(p^\mu)
\end{array}\right)\quad.
\end{equation}
Obviously the kinematic structure derived from these spinors, such as the
wave equation, the field operator, and transformation of the physical
states under Parity, Charge Conjugation, and Time Reversal,  would be very
different from those already discussed. Parity is non-maximally violated
for $0\, <\,\theta\,<\,\pi/2$. Of all the possible kinematic structures,
which of these are physically realized in nature depends on  what
symmetries are respected in Nature and to what extent. Once this question
is answered by experimental observations, one may proceed to seek an
appropriate kinematic structure to build the interacting theory.

Even though we have not investigated all possible physically relevant
requirements that may be used to fix relative phase factors between the
$(j,\,0)$ and $(0,\,j)$ spinors our work has been presented in sufficient
detail to reduce all such generalizations to conceptually simple and
algebraically well defined exercise. It is hoped that such generalizations
will be considered in future as necessitated by specific problems at hand.

In summary,
we argue that  constraints imposed by the kinematic structure
on the dynamical aspects of a theory are subtler than the textbook
treatment of this subject would indicate. The kinematic symmetries relevant
to the problem determine the underlying kinematic structure on which to
build one's dynamical theory. Based on the observation that if
$\phi_{_{L}}(p^\mu)$ transforms as a $(0,\,j)$ spinor under Lorentz boosts,
then $\Theta_{[j]}\,\phi_{_{L}}^\ast(p^\mu)$ transforms as a $(j,\,0)$
spinor (with a similar relationship existing between $\phi_{_{R}}(p^\mu)$
and $\Theta_{[j]}\,\phi_{_{R}}^\ast(p^\mu)$ we introduced McLennan-Case
type $(j,\,0)\oplus(0,\,j)$ spinors. Relative phases between
$\phi_{_{R}}(p^\mu)$ and $\Theta_{[j]}\,\phi_{_{R}}^\ast(p^\mu)$, and
$\Theta_{[j]}\,\phi_{_{L}}^\ast(p^\mu)$ and $\phi_{_{L}}(p^\mu)$, turn out
to have physical significance and are fixed by appropriate requirements.
Explicit construction, and a series of physically relevant properties, for
these spinors were obtained for spin-$1/2$ and spin-$1$ culminating in the
construction of a fundamentally new wave equation and introduction of
Dirac-like and Majorana-like quantum fields.

\vskip 0.5in {\it Acknowledgements}
Apart from the acknowledgements expressed at various places in the text I
wish to extend my {\it zimpoic} thanks to Christoph Burgard for a series of
insightful conversations on some of the subject matter of this manuscript.
In addition, I thank the following colleagues and friends at the Los Alamos
National Laboratory for their constant accessibility and encouragement:
George Glass, Terry Goldman, Steve Greene, Peter Herczeg, Cy Hoffman,
Mikkel Johnson, Gus Sinnis, and Hywel White; and in particular, all members
of the LSND (experimental group studying neutrino oscillations at LAMPF)
and MILAGRO (experimental group studying gamma-ray bursts at LANL) for
educating and entertaining me with the various mysteries of the Universe
during the course of this work. Away from the Laboratory, I enjoyed many
conversations on the subject with  Rabindra N. Mohapatra at Snowmass '94,
and Professor Pierre Ramond kindly directed me to an important paper in the
field. I thank them both. The work was indeed inspired by the comment on
``magic of the Pauli matrices'' (which turned out to be ``magic of Wigner's
$\Theta_{[j]}$ operator'' as we saw in this work) that Professor Pierre
Ramond made in his book (Ref. \cite{PR}, p. 16). Professor Steven Weinberg
kindly  commented \cite{SWpc} on why Ref. \cite{FNBWW} gets ``such  a
surprising result,'' I wish to thank him for sharing his thoughts on the
subject. I wish to thank Jeanne Bowles
for a careful reading of the manuscript and for her
suggestions on stylistic  matters.

{\it This work was done under the auspices of the U. S. Department of
Energy.}

\begin{table} \caption{For spin-$1/2$,  covariant norms for
$\lambda^S(p^\mu)$ and $\lambda^A(p^\mu)$ for general values of
$\vartheta_1$ and $\vartheta_2$. The value of
$\overline{\lambda}_{\eta^\prime}(p^\mu)\,
\lambda_{\eta^{\prime\prime}}(p^\mu)$ is tabulated at the intersection of
appropriate row and column. }\tablenotemark[1]\tablenotetext[1] {For the
choice of phases $\vartheta_1\,+\,\vartheta_2\,=\,0$, we obtain a
bi-orthonormal set with sub-spaces spanned by $\lambda^S(p^\mu)$ and
$\lambda^A(p^\mu)$ mutually orthogonal. } \begin{tabular}{ccccc}
\multicolumn{1}{c}{} & \multicolumn{1}{c}{$\lambda^S_\uparrow(p^\mu)$} &
\multicolumn{1}{c}{$\lambda^S_\downarrow(p^\mu)$}
&\multicolumn{1}{c}{$\lambda^A_\uparrow(p^\mu)$} &
\multicolumn{1}{c}{$\lambda^A_\downarrow(p^\mu)$}\\ \tableline
$\overline{\lambda}^S_\uparrow(p^\mu)\,\,\,$& $0\,\,$ & $
2\,i\,N^2\,\cos(\vartheta_1+\vartheta_2) \,\,$&$ 0\,\,$&$
-\,2\,N^2\,\sin(\vartheta_1+\vartheta_2)\,\,$\\
$\overline{\lambda}^S_\downarrow(p^\mu)\,\,$&$
-\,2\,i\,N^2\,\cos(\vartheta_1+\vartheta_2) \,\,$&$ 0 \,\,$&$
2\,N^2\,\sin(\vartheta_1+\vartheta_2) \,\,$&$0\,\,$\\
$\overline{\lambda}^A_\uparrow(p^\mu)\,\,$&$
0\,\,$ & $2\,N^2\,\sin(\vartheta_1+\vartheta_2)\,\,$&$ 0 \,\,$&$
-\,2\,i\,N^2\,\cos(\vartheta_1+\vartheta_2)\,\,$ \\
$\overline{\lambda}^A_\downarrow(p^\mu)\,\,$&$
-\,2\,N^2\,\sin(\vartheta_1+\vartheta_2)\,\,$&$ 0 \,\,$&$
2\,i\,N^2\,\cos(\vartheta_1+\vartheta_2)
\,\,$&$0\,\,$\\
\end{tabular}
\end{table}

\begin{table}
\caption{For spin-$1$,  covariant norms for $\lambda^S(p^\mu)$ and
$\lambda^A(p^\mu)$ for general values of $\delta_1$, $\delta_2$, and
$\delta_3$. The value of $\overline{\lambda}_{\eta^\prime}(p^\mu)\,
\lambda_{\eta^{\prime\prime}}(p^\mu)$ is tabulated at the intersection of
appropriate row and column.
}\tablenotemark[1]\tablenotetext[1]{Here we have introduced the following
abbreviations: $\alpha_\pm\,=\,{\rm N}^2
\exp{\bbox{(}}-\,i\,(\delta_3\,+\,\delta_1){\bbox)}
\,\left[\exp{\bbox{(}}2\,i\,(\delta_3\,+\,\delta_1){\bbox)} \,\pm\,1\right]
$ and $ \beta_\pm\,=\,{\rm N}^2
\exp(-\,2\,i\,\delta_2)\,\left[\exp(4\,i\,\delta_2)\,\pm\,1\right]$.
Equivalently, $\alpha_+\,=\,2\,{\rm N}^2\,\cos(\delta_3\,+\delta_1)$,
$\alpha_-\,=\,2\,i\,{\rm N}^2\,\sin(\delta_3\,+\delta_1)$, $\beta_+\,=\,
2\,{\rm N}^2\,\cos(2\,\delta_2)$, and $\beta_-\,=\,
2\,i\,{\rm N}^2\,\sin(2\,\delta_2)$. For the choice of phases
$\delta_3\,+\,\delta_1\,=\,0$ and $\delta_2\,=\,0$, we obtain a bi-orthonormal
set with sub-spaces spanned by $\lambda^S(p^\mu)$ and $\lambda^A(p^\mu)$
mutually orthogonal.}
\begin{tabular}{ccccccc}
\multicolumn{1}{c}{} & \multicolumn{1}{c}{$\lambda^S_\uparrow(p^\mu)$} &
\multicolumn{1}{c}{$\lambda^S_\rightarrow(p^\mu)$} &
\multicolumn{1}{c}{$\lambda^S_\downarrow(p^\mu)$}
&\multicolumn{1}{c}{$\lambda^A_\uparrow(p^\mu)$} &
\multicolumn{1}{c}{$\lambda^A_\rightarrow(p^\mu)$} &
\multicolumn{1}{c}{$\lambda^A_\downarrow(p^\mu)$}\\
\tableline
$\overline{\lambda}^S_\uparrow(p^\mu)\,\,\,$&
$0$ & $0$ & $\alpha_+$ & $0$ & $0$ & $\alpha_-$\\
$\overline{\lambda}^S_\rightarrow(p^\mu)\,\,\,$&
$0$ & $-\,\beta_+$ & $0$ & $0$ & $-\,\beta_-$ & $0$ \\
$\overline{\lambda}^S_\downarrow(p^\mu)\,\,$&
$\alpha_+$ & $0$ & $0$ & $\alpha_-$ & $0$ & $0$\\
$\overline{\lambda}^A_\uparrow(p^\mu)\,\,$&
$0$ & $0$ & $-\,\alpha_-$ & $0$ & $0$ & $-\,\alpha_+$ \\
$\overline{\lambda}^A_\rightarrow(p^\mu)\,\,\,$&
$0$ & $\beta_-$ & $0$ & $0$ & $\beta_+$ & $0$ \\
$\overline{\lambda}^A_\downarrow(p^\mu)\,\,$&
$-\,\alpha_-$ & $0$ & $0$ & $-\,\alpha_+$ & $0$ & $0$ \\
\end{tabular}
\end{table}

\end{document}